\documentstyle[11pt]{article}
\newcommand{\be}{\begin{equation}}
\newcommand{\ee}{\end{equation}}

\begin{document}

\begin{titlepage}

\vskip 2. cm
\begin{center}
\vfill
{\large\bf  Note on the Time Reversal Asymmetry \break
of Equations of Motion\footnote{\ Unfortunately, several
misleading errors have been edited into the
printed version of this contribution (Foundations of Physics
Letters, {\bf 12}, 193), including its title. Note that this
e-print represents the correct version.} }
\vskip 2.6cm

{\bf H. D. Zeh}
\vskip 0.4cm
Institut f\"ur Theoretische Physik, Universit\"at Heidelberg,\\
Philosophenweg 19, D-69120 Heidelberg, Germany.\\
e-mail: zeh@urz.uni-heidelberg.de
\end{center}

\vskip 2 cm
\begin{center}
{\bf Abstract}
\end{center}
\begin{quote}
Rohrlich's recent claim that the equation of motion for a point
charge be symmetric under time reversal is shown to be the
result of an unusual definition. The equation of motion for
a charged sphere of finite size, which in contrast is claimed to
be asymmetric because of the finite propagation time of its
(retarded) self-forces, is shown to possess the same
asymmetry (or the same symmetry, depending on the definition) as
that for a point charge. Similar arguments apply to other
effective equations of motion (such as those describing friction or
decoherence).

\vskip .5cm

 \end{quote}

\vskip 1. cm

\end{titlepage}

Rohrlich argued recently\cite{1} that the
Lorentz-(Abra\-ham)-Dirac (LAD) equation for a point charge moving
with four-velocity $v^\mu(\tau)$  along its world line,
$$
m\dot v^\mu = {2e^2\over 3} (\ddot v^\mu - v^\mu \dot v^\nu \dot
v_\nu) + F^{\mu\nu}_{\rm in} v_\nu \quad , \eqno(1)
$$
be invariant under time reversal. Here, $m$ is the renormalized
(physical) mass, while $F^{\mu\nu}_{\rm in}$  is an external
Maxwell field. Rohrlich's claim may appear surprising, since this
equation contains terms proportional to the velocity and to the
first time derivative of the acceleration, both of which change
sign under time reversal --- similar to the terms responsible for
friction in the equation of motion for a mass point. These two
terms describe the {\it loss} of energy that must balance the
{\it emission} of radiation (Dirac's radiation reaction) and the
ill-defined self-acceleration that leads to exponentially
increasing velocity. Both terms arise from the presumed
retardation (the absence of the advanced fields of the point
charge).

It may appear even
more surprising when Rohrlich also insists that the
Caldi\-rola-Yaghjian (CY) equation\cite{2,3} for a charged sphere
of radius $a$,
$$
m_0 \dot v^\mu (\tau) = {2e^2 \over 3a} {v^\mu (\tau - 2a) + v^\mu
(\tau) v^\nu (\tau) v_\nu (\tau - 2a) \over 2a} + F^{\mu\nu}_{\rm
in} v_\nu (\tau) \quad ,
\eqno(2)
$$
with bare mass $m_0 = m - 2e^2/{3a}$, be asymmetric, even though
the former equation can be obtained from the latter in the limit
$a \to 0$  under a time-symmetric though divergent mass
renormalization. Apparently, the idea underlying these claims is
that only the retarded self-forces within the sphere create an
asymmetry, which must then disappear in the point limit. I will
now argue that this picture is wrong.

	What Rohrlich does in fact prove in the first part of his paper
 (and also in his book\cite{4}) is the invariance of the LAD
equation under time reversal provided this is defined to include a
simultaneous interchange of incoming and outgoing fields. However,
this symmetry of the {\it global} situation merely reflects that
of the complete theory; it does not represent the behavior under
time reversal of a point charge that would always be allowed
freely to create its {\it retarded} radiation. Although the claim
is then formally correct, it is based on a very unusual and
misleading definition of time reversal for an equation of motion.
This becomes obvious when the definition is correspondingly
applied to friction, which would also be time reversal invariant
if the second law were simultaneously reversed (that is, if
dissipated heat were replaced with heat focussing on
the mass point instead of the retarded fields being replaced with
advanced ones in the case of the charge). In the same sense, the
Lorentz force is symmetric under time reversal if its magnetic
fields are simultaneously reflected in space. These three
situations differ only in the complexity of their ``environments",
and hence in the {\it practical} difficulties of time-reversing
them for this Loschmidt-type argument. Their symmetry is thus
trivial (as Rohrlich correctly points out for the LAD equation),
but essentially of mere theoretical importance in the first two
cases.

	Rohrlich's symmetry of the LAD equation is based on the general
equivalence of the two familiar representations of an arbitrary
field, {\it viz.} either as a sum of incoming and retarded fields
(of the sources in the considered spacetime volume) or of
outgoing and advanced fields. Their essential difference is
that it is easy to control incoming fields, but hard to
manipulate outgoing ones. Moreover, fields often vanish
before sources are turned on, while they do not after the
sources are turned off. This familiar {\it fact}  is a
consequence of the general presence of absorbers (such as
laboratory walls) with their thermodynamical arrow of
time\cite{5}.

This equivalence of different representations may also be applied
to the nonsingular case of a charged sphere of finite size. One
may either add its retarded field to a given incoming field or the
advanced field to the outgoing one in order to satisfy the Maxwell
equations. The first choice leads to the CY equation (2),
while the second one would mean that the retarded arguments
$\tau - 2a$ in (2) have to be replaced with $\tau + 2a$  because
of the advanced self-forces that are now {\it required} to act
within the charged sphere. Instead of this replacement, Rohrlich
leaves this expression in its retarded form in spite of his
definition of time reversal, since ``advanced interactions are
never observed" and ``should not be possible''. However, the
interchange of incoming and outgoing fields required in his
definition of time reversal should then neither be possible. (In
practice, one would have to prepare the complete though
time-reversed retarded field coherently as an incoming field.)
Retardation or advancement are here a consequence of the chosen
representation --- not of the empirical situation in our world.
When formally fixing outgoing fields, one has to use advanced
fields of the considered sources everywhere. Advanced external
fields would be inconsistent in conjunction with retarded
internal fields.

One may then consider the limit $a \to 0$  in the CY equation by
using the Taylor expansion $v^\mu (\tau \mp 2a) = v^\mu (\tau)
\mp 2a \dot v^\mu (\tau) + 2a^2 \ddot v^\mu(\tau) + \dots$  and
the condition  $v^\mu v_\mu = -1$ together with its time
derivatives (that is,  $v^\mu \dot v_\mu = 0$ and  $v^\mu \ddot
v_\mu = -\dot v^\mu \dot v_\mu$). The first order of this Taylor
expansion gives the mass renormalization  $\Delta m = 3e^2/2a$,
while the second one leads precisely to the LAD equation, with
signs of the retardation differing in the two cases.
The limit  $a \to +0$ is thus nontrivially different from the
limit    $a \to -0$. (This is related to the well known fact that
master equations are trivial in first order of the
interaction.) Therefore, the LAD equation and the CY equation
possess the {\it same asymmetry} under time reversal in the usual
sense, and the {\it same symmetry} in the sense of Rohrlich.

Equivalent concepts of time reversal are valid for other
equations of motion\cite{6}. They can also be applied to the
master equation of a mass point described quantum mechanically
under the effect of decoherence\cite{7,8},
$$
i{\partial \rho (x,x^\prime,t) \over \partial t} = {1\over
2m}\left( {\partial^2 \over \partial {x^\prime} ^2 } - {\partial^2
\over \partial x^2 } \right) \rho -i\lambda (x - x^\prime)^2 \rho
\quad ,
\eqno(3)
$$
where $\rho$ is the density matrix. This equation is again
asymmetric in the usual sense,  since it reflects the {\it
formation of retarded entanglement} (quantum correlations). {\it
All} these equations of motion are asymmetric if regarded by
themselves, since their dynamical objects (such as mass points)
are strongly coupled to a time-directed environment, while they
would be symmetric if they were time-reversed together with their
environment.

	In the case of decoherence, time reversal  of the environment
would require recoherence, that is, the conspiratorial presence
of previously unobserved but presicely matching Everett
components (``parallel worlds"). However, a fundamental collapse
of the wave function (if it existed as a dynamical law) would not
even theoretically allow the environment to be time-reversed.
While all these situations may reflect the same master arrow of
time\cite{5} (that is, the same cosmic initial condition), it
remains open whether there is a boundary somewhere that
separates reversible from irreversible
physics in
a fundamental (law-like) way. Most physicists appear ready to
accept such a boundary in conjunction with equation (3), that is,
for quantum mechanical measurements or related ``probabilistic
quantum events" --- wherever the precise boundary between this
collapse of the wave function and the realm of the Schr\"odinger
equation may be located.

 \eject


\vspace{10mm}

\begin{thebibliography}{99}
\bibitem{1}
  F. Rohrlich, Found. Phys. {\bf 28}, 1045 (1998).
\bibitem{2}
 P. Caldirola, Nuovo Cim. {\bf 3}, Supp. 2, 297 (1956).
\bibitem{3}
A.D. Yaghjian, {\it Relativistic Dynamics of a Charged Sphere}
(Springer, Berlin, 1992).
\bibitem{4}
 F. Rohrlich, {\it Classical Charged Particles} (Addison-Wesley,
Reading, 1965).
\bibitem{5}
 H. D. Zeh, {\it The Physical Basis of the Direction of Time}, 3rd
edn. (Springer, Berlin, 1999).
\bibitem{6}
 T.C. Quinn and R.M. Wald, Phys. Rev. {\bf D56}, 3381 (1997).
\bibitem{7}
 D. Giulini, E. Joos, C. Kiefer, J. Kupsch, I.-O. Stamatescu, and
H.D. Zeh, {\it Decoherence and the Appearance of a Classical World
in Quantum Theory} (Springer, Berlin, 1996).
\bibitem{8}
R. Omn\`es, Phys. Rev. {\bf A56}, 3383 (1997).


\end{thebibliography}
\end{document}